\documentclass[12pt,twoside,a4paper]{article}
\usepackage[dvips]{epsfig}
\newcommand{\simgt}{\,\rlap{\lower 3.5 pt \hbox{$\mathchar \sim$}}
  \raise 1pt \hbox {$>$}\,}
\newcommand{\simlt}{\,\rlap{\lower 3.5 pt \hbox{$\mathchar \sim$}}
  \raise 1pt \hbox {$<$}\,}

\begin{document}
\thispagestyle{empty} 
\title{
\vskip-3cm
{\baselineskip14pt
\centerline{\normalsize DESY 99--035 \hfill ISSN 0418--9833}
\centerline{\normalsize MZ-TH/99--05 \hfill} 
\centerline{\normalsize TTP99-13 \hfill}
\centerline{\normalsize hep--ph/9903377 \hfill} 
\centerline{\normalsize March 1999 \hfill}} 
\vskip1.5cm
Photon Plus Jet Production in Large-$Q^2$ $ep$ Collisions \\
at Next-to-Leading Order QCD 
\author{A.~Gehrmann-De Ridder$^1$, G.~Kramer$^2$ and H.~Spiesberger$^3$
\vspace{2mm} \\
{\normalsize $^1$ Institut f\"ur Theoretische Teilchenphysik,
  Universit\"at Karlsruhe,}\\ 
{\normalsize D-76128 Karlsruhe, Germany} \vspace{2mm}\\
{\normalsize $^2$ II. Institut f\"ur Theoretische
  Physik\thanks{Supported by Bundesministerium f\"ur Forschung und
    Technologie, Bonn, Germany, under Contract 05~7~HH~92P~(0), and
    by EU Fourth Framework Program {\it Training and Mobility of
    Researchers} through Network {\it Quantum Chromodynamics and
    Deep Structure of Elementary Particles}
    under Contract FMRX--CT98--0194 (DG12 MIHT).}, Universit\"at
  Hamburg,}\\ 
\normalsize{Luruper Chaussee 149, D-22761 Hamburg, Germany} \vspace{2mm}
\\ 
\normalsize{$^3$ Institut f\"ur Physik,
  Johannes-Gutenberg-Universit\"at,}\\ 
\normalsize{Staudinger Weg 7, D-55099 Mainz, Germany} \vspace{2mm} \\
\normalsize{e-mail: gehra@particle.physik.uni-karlsruhe.de,} \\
\normalsize{kramer@mail.desy.de, hspiesb@thep.physik.uni-mainz.de}
} }

\date{}
\maketitle
\begin{abstract}
\medskip
\noindent
The production of photons accompanied by jets in large-$Q^2$ deep
inelastic $ep$ scattering is calculated in next-to-leading order. We
describe how to consistently include contributions from quark-to-photon
fragmentation and give numerical results relevant for HERA experiments.
\end{abstract}

\section{Introduction}

The production of isolated photons in high-energy hadronic collisions is
an important testing ground for QCD. Since the photon does not take part
in the strong interaction, it is a "direct" probe of the hard scattering
process and provides a means to measure the strong coupling constant
$\alpha_s$ or to extract information on the parton distributions, in
particular the gluon density in the proton \cite{1}. Moreover, good
knowledge of the standard model predictions for direct photon production
is required since it is an important background for many searches of new
physics.

At HERA, with increasing luminosity, the measurement of isolated photon
production will give information on the parton content of the proton and
at $Q^2=0$, i.e. for photoproduction, also on the parton distributions
in the photon. First experimental results from the ZEUS collaboration
\cite{2,3} (see also \cite{3a}) have been reported and found in
reasonable agreement with next-to-leading order (NLO) predictions
\cite{4}. Cross sections for the production of hard photons in deep
inelastic scattering are much smaller as compared to photoproduction
and, therefore, more difficult to measure. Typical cross sections, for
example with $Q^2 > 10~GeV^2$, are of the order of $10~pb$. With a
luminosity of $50~pb^{-1}$ one thus expects measurements of differential
cross sections to become feasible.

A NLO calculation for direct photon production in deep inelastic $ep$
scattering, $ep \rightarrow e\gamma X$, at large $Q^2$ has been reported
recently by two of us and D. Michelsen \cite{5}. Since hard photon
production occurs, compared to inclusive deep inelastic scattering, at a
relative order O($\alpha$) one expects a sizable cross section only at
moderately large $Q^2$. Therefore, one can restrict the calculation to
pure virtual photon exchange with $Z$ exchange neglected. In \cite{5}
the hadronic final state is separated into $\gamma + (1+1)$- and $\gamma
+ (2+1)$-jet topologies (the remnant jet being counted as "+1" jet us
usual). The approach is thus analogous to the calculation of $(2+1)$-
and $(3+1)$-jet cross sections, where one of the final state gluons is
replaced by a photon \cite{6}.  In addition to the direct production,
photons can also be produced through the fragmentation of a hadronic jet
into a single photon carrying a large fraction of the jet energy
\cite{7}. This long-distance process is described in terms of the
quark-to-photon and gluon-to-photon fragmentation functions which absorb
collinear singularities present in the perturbative calculation. First
measurements of the $q \rightarrow \gamma $ fragmentation function in
$e^+e^- \rightarrow \gamma + 1$-jet are presented in \cite{8} (see also
\cite{9} for the discussion of an inclusive measurement). The NLO theory
for this process has been worked out in \cite{10}. In \cite{5} the
fragmentation contributions were discarded and the photon-quark
collinear singularities had been removed by explicit parton-level
cutoffs. The results depended strongly on these photon-parton cutoffs,
in particular for the incoming gluon contributions \cite{5}. However,
these cutoffs are difficult to control experimentally.

In this paper we report results in which the fragmentation contributions
are included together with isolation criteria that limit the hadronic
energy in the jet containing the photon. Whereas in \cite{5} various
photon + jet cross sections with invariant mass jet resolution criteria
were calculated, in this work we concentrate on the calculation of
various differential cross sections either exclusively or inclusively
which depend on the transverse momenta and rapidities of the photon or
the accompanying jet.  We use the $\gamma^{*} p$ center-of-mass system
to define the kinematic variables. The cone method is applied to define
the parton jets and to isolate the photon signal.

In section 2 a brief outline of the theoretical background for
calculating the cross section is given. In section 3 numerical results
are presented.  Section 4 contains a short summary and the conclusions.

\section{Subprocesses Through Next-to-Leading Order}

\subsection{Leading-Order Contributions}

In leading order, the production of photons in deep inelastic electron
(positron) scattering is described by the quark (antiquark) subprocess
\begin{equation}
   e(p_1)+q(p_3) \rightarrow e(p_2)+q(p_4)+\gamma (p_5)
\end{equation}
where the particle momenta are given in parentheses. The momentum of the
incoming quark is a fraction $\xi$ of the proton momentum $p_P$,
$p_3=\xi p_P$.  The proton remnant $r$ has the momentum
$p_r=(1-\xi)p_P$. It hadronizes into the remnant jet so that the process
(1) gives rise to $\gamma + (1+1)$-jet final states. In the virtual
photon $\gamma ^{*}$-proton center-of-mass system the hard photon
recoils against the hard jet back-to-back. To remove photon production
by incoming photons $\gamma ^{*}$ with small virtuality (photoproduction
channel) and to restrict to the case where the scattered electron
$e(p_2)$ is observed, one applies cuts on the usual deep inelastic
scattering variables $x,~y$ and $Q^2$. In addition, to have photons
$\gamma (p_5)$ of sufficient energy we require an explicit cut on the
invariant mass $W$ of the final state, $W^2=(q+p_P)^2$, where $q$ is the
electron momentum transfer, $q=p_1-p_2$ and $Q^2=-q^2$ as usual.  Both 
leptons and quarks emit photons. The subset of diagrams where the
photon is emitted from the initial or final state lepton (leptonic
radiation) is explicitly gauge invariant and can be considered
separately. Similarly, the diagrams with a photon emitted from quark
lines is called quarkonic radiation. In addition, there are also
contributions from the interference of these two. Since we are
interested in testing QCD under the circumstances that the photon is
emitted from quarks the contributions from leptonic radiation are viewed
as a background and must be suppressed. This can easily be done by a cut
on the photon emission angle with respect to the incoming electron 
\cite{5}. In our numerical evaluation we include this background source 
as well as the interference contribution.

At lowest order, each parton is identified with a jet and the photon is
automatically isolated from the quark jet by requiring a non-zero
transverse momentum of the photon or jet in the $\gamma ^{*}$-proton
center-of-mass frame. Therefore the photon fragmentation contribution is
absent at this order.

\subsection{Next-To-Leading Order Corrections}

At NLO, processes with an additional gluon, either in the final state or
in the initial state, must be taken into account, i.e.

\begin{eqnarray}
 e(p_1)+q(p_3) \rightarrow e(p_2)+q(p_4)+\gamma (p_5)+g(p_6), 
\\
 e(p_1)+g(p_3) \rightarrow e(p_2)+q(p_4)+\gamma (p_5)+\bar{q}(p_6), 
\end{eqnarray}

where the momenta of the particles are again given in parentheses. In
addition, virtual corrections (one-loop diagrams at O($\alpha_s$)) to
the LO process (1) have to be included. The complete matrix elements for
(2) and (3) are given in \cite{11}. The processes (2) and (3) contribute
both to the $\gamma + (1+1)$-jets cross section, as well as to the cross
section for $\gamma +(2+1)$-jets. In the latter case each parton in the
final state of (2) and (3) builds a jet on its own, whereas for $\gamma
+(1+1)$-jets a pair of final state partons is experimentally unresolved.
The exact criteria for combining two partons into one jet will be
introduced when we present our results. Following the customary
experimental procedure the resolution constraints will be based on the
cone algorithm.

In the calculation of the cross section for $\gamma + (1+1)$-jets we
encounter the well-known infrared and collinear singularities. They
appear for the processes (2) and (3) in those phase space regions where
two partons are degenerate to one parton, i.e. when one of the partons
becomes soft or two partons become collinear to each other. The
singularities are assigned either to the initial state (ISR) or to the
final state (FSR).  Contributions involving the product of an ISR and a
FSR factor are separated by partial fractioning. The FSR singularities
cancel against singularities from the virtual corrections to the LO
process (1). For the ISR singularities, this cancellation is incomplete
and the remaining singular contributions have to be factorized and
absorbed into the renormalized parton distribution functions (PDF's) of
the proton.

To accomplish this procedure, the singularities are isolated in an
analytic calculation with the help of dimensional regularization. This
is difficult for the complete cross sections of the processes (2) and
(3). After partial fractioning, the phase-space slicing method \cite{12}
is used to separate the singular regions in the 4-particle phase space
and to determine in these regions the approximated matrix elements and
phase space factors. In those regions only, the calculation is performed
analytically. For this purpose a slicing cut $y_0^J$ is applied to the
scaled invariant masses $y_{ij}$, where $y_{ij}=(p_i+p_j)^2/W^2$ with
$W^2=(p_P+p_1-p_2-p_5)^2$.  $y_0^J$ must be chosen small enough, so that
terms of the order O($y_0^J$), which are discarded due to the singular
approximation, are so small that an accuracy of a few percent can be
achieved for the final result.

To be somewhat more explicit, let us assume that by partial fractioning
the contribution proportional to the pole term $1/y_{46}$ has been
isolated in the matrix element $|M|^2$ for the process (2). In the
infrared region $p_6 \rightarrow 0$ the two invariants $y_{46}$ and
$y_{36}$ vanish.  Then the integration over these two variables is
performed: ($i$) over the singular region (S), $y_{46}<y_0^J$, $y_{36}
>0$, analytically with $4-2\epsilon$ dimensions and using the singular
approximation; ($ii$) over the finite region (F), $y_{46} > y_0^J$,
$y_{36} < y_0^J$, numerically without any approximation to $|M|^2$ and
in 4 dimensions; and ($iii$) over the explicit two-parton region (R),
$y_{46} > y_0^J$, $y_{36} > y_0^J$, also numerically.  This separation
yields two contributions, the parton-level $\gamma + (1+1)$-jets, which
come from the integration over the regions (S) and (F) and the
parton-level $\gamma + (2+1)$-jets contribution, which corresponds to
the integration over the region (R). All the remaining phase space
integrations are performed numerically with usual Monte Carlo routines.
For the physical cross sections defined in the next section, which are
obtained by adding the contributions from the regions S, F and R, of
course, and after adding the virtual contributions and performing the
subtraction of the remaining ISR collinear singularities, the dependence
on the slicing parameter $y_0^J$ cancels. This has been checked in
\cite{5}. This means that the cut-off $y_0^J$ is purely technical.

In addition, the matrix elements $|M|^2$ for the processes (2) and (3)
have {\em photonic} infrared and collinear singularities, i.e.\ due to
soft or collinear photons. The infrared singularity can not occur since
we require a sufficiently large photon energy $E_{\gamma}=|\vec{p_5}|$.
But collinear singularities are present in general.  In the earlier work
\cite{5} these $q$-$\gamma$ collinear contributions were eliminated by
an isolation cut on the photon of the form $y_{5i}>y_0^{\gamma}$ with a
sufficiently large isolation parameter $y_0^{\gamma}$ which was
considered as a physical cut. In this approach, the photon was
considered as a special parton, which was always isolated from all other
partons in the initial and final state. Such a photon isolation is very
difficult to impose experimentally, since it refers to a separation of
the photon from partons, whereas in the experiment only hadrons are
measured directly which are recombined to jets. In addition, it was
found in \cite{5} that the results, in particular for the
gluon-initiated process (3), depend strongly on the isolation cut
$y_0^{\gamma}$. In (3) the photon can become collinear to two final
state partons, namely $q$ and $\bar{q}$, which explains the stronger
dependence compared to the process (2), where there is only one quark
(or antiquark) in the final state. Although under the kinematical
conditions assumed in \cite{5} the contribution of the process (3) was
only some fraction of the total cross section for $\gamma +(1+1)$-jets,
the $y_0^{\gamma}$ dependence of the final result was undesirable. In a
more systematic treatment the $y_0^{\gamma}$ dependence can be avoided
by adding contributions from the quark-to-photon fragmentation function
(FF). In order to achieve this we include the contributions to $|M|^2$
from (2) and (3), where $y_{5i} < y_0^{\gamma}$ with $i = 4$ in (2) and
$i = 4$ and 6 in (3). This leads to collinear divergent contributions
which are regulated by dimensional regularization. The divergent part is
absorbed into the bare photon FF to yield the renormalized FF denoted by
$D_{q \rightarrow \gamma}$. The additional fragmentation contribution,
which includes the contributions from the region $y_{5i} < y_0^{\gamma}$
($i=3,4$), has, for example, for the process (2) the following form
\begin{eqnarray}
  \mid M\mid^2_{\gamma^{*}q \rightarrow qg\gamma} =
 \mid M\mid^2_{\gamma^{*}q\rightarrow qg} \otimes D_{q\rightarrow\gamma}(z)
\end{eqnarray}
where $D_{q \rightarrow \gamma}(z)$ in (4) is given by \cite{13}
\begin{eqnarray}
 D_{q \rightarrow \gamma}(z) = D_{q \rightarrow \gamma}(z,\mu_F^2) +
 \frac{\alpha e_q^2}{2\pi}
  \left(P_{q\gamma}(z)\ln\frac{z(1-z)y_0^{\gamma}W^2}{\mu_F^2} +
    z\right). 
\end{eqnarray}
$P_{q\gamma}(z)$ is the LO quark-to-photon splitting function 
\begin{equation}
    P_{q\gamma}(z) = \frac{1+(1-z)^2}{z}
\end{equation}

and $e_q$ is the electric quark charge. $D_{q \rightarrow
  \gamma}(z,\mu_F^2)$ stands for the non-perturbative FF of the
transition $q \rightarrow \gamma $ at the factorization scale $\mu_F $.
This function will be specified in the next section when we present our
results. The second term in (5), if substituted in (4), is the finite
part of the result of the integration over the collinear region $y_{5i}
< y_0^{\gamma }$.  As will be explicitly shown in \cite{GKS}, the
$y_0^{\gamma }$ dependence in (5) cancels the dependence of the
numerically evaluated $\gamma +(1+1)$-jet cross section restricted to
the region $y_{5i} > y_0^{\gamma }$, investigated in \cite{5}.  The
variable $z$ stands for the fraction of the photon energy in terms of
the energy of the quark that emits the photon. Suppose the photon is
emitted from the final state quark with 4-momentum $p_4'=p_4+p_5$, then
$z$ can be related to the invariants $y_{35}$ and $y_{34}$
\begin{equation}
 z= \frac{y_{35}}{y_{34'}} = \frac{y_{35}}{y_{34}+y_{35}}
\end{equation}

The fragmentation contribution is proportional to the cross section for
$\gamma^{*} q \rightarrow qg$, which is of O($\alpha_s$) and well known.
It must be convoluted with the function in (5) as indicated in (4) to
obtain the contribution to the cross section for $\gamma^{*} q
\rightarrow qg\gamma $ at $O(\alpha \alpha_s)$.  Equivalent
formulas are used to calculate the fragmentation contributions to the
channel (3) and in the case where the quark in the initial and final
state is replaced by an antiquark in (2).

\section{Results}

The results presented in this section are obtained for energies and
kinematical cuts appropriate for the HERA experiments. The energies of
the incoming electron (positron) and proton are $E_e=27.5~GeV$ and
$E_P=820~GeV$, respectively. The cuts on the usual DIS variables are
\begin{eqnarray}
   Q^2 \geq 10~GeV^2,~~~~~~W>10~GeV,\\ \nonumber
   10^{-4} \leq x \leq 0.5,~~~~~~0.05 \leq y \leq 0.95.
\end{eqnarray}
To eliminate the background from lepton radiation \cite{5} we require 
\begin{eqnarray}
    90^{\circ} < \theta_{\gamma} < 173^{\circ},~~~ 
    \theta_{\gamma e} \geq 10^{\circ} 
\end{eqnarray}
where $\theta_{\gamma}$ is the emission angle of the photon measured
with respect to the momentum of the incoming electron in the HERA
laboratory frame.  The cut on $\theta_{\gamma e} = \,\,\rlap)\!\!\!<
(e(p_2), \gamma(p_5))$ suppresses leptonic radiation from the
final-state electron. The PDF's of the proton are taken from \cite{14}
(MRST) and $\alpha_s$ is calculated from the two-loop formula with the
same $\Lambda $ value ($\Lambda_{\overline{MS}}(n_f=4) = 300~MeV$) as 
used in the MRST parametrization of the proton PDF.  The scale in 
$\alpha_s$ and the fractorization scale are equal and fixed to $Q^2$.

We are interested in the differential two-particle inclusive cross
section $E_{\gamma}E_J d\sigma/d^3p_{\gamma}d^3p_J$ at NLO (up to 
$O(\alpha \alpha_s)$), where $(E,\vec p)$ represents the four-vector
momentum of the $\gamma $ or jet.  The $\gamma + (1+1)$-jet cross 
section receives contributions from leading and next-to-leading order 
and the $\gamma + (2+1)$-jet cross section from leading order only. In 
the latter case, only $\gamma + 2$-parton-level jets contribute, while 
each parton including the photon build a jet on their own.  The 
evaluation of the $\gamma + (1+1)$-jet cross section is based on two 
separate contributions, a set of two-body contributions, i.e. $\gamma + 
1$ parton-level jet, and a set of three-body contributions, i.e. $\gamma 
+ 2$ parton-level jets.  In this definition of parton-level jets the
remnant jet $"+1"$ is not counted whereas the photon is considered also
as a parton, like $q,\bar{q}$ and $g$. Each set of contributions is
completely finite as all infrared and collinear singularities have been
canceled or absorbed into the proton PDF or the quark-to-photon FF. Each
contribution to the $\gamma + (1+1)$-jet cross section depends
separately on the slicing parameter $y_0^J$.  The analytic contributions
are valid only for very small $y_0^J$. Separately, the two contributions
have no physical meaning. For the contributions to the $\gamma +
(1+1)$-jet cross section coming from $\gamma +2$ parton-level jets, a
slicing cut $y_0^{\gamma}$ for the photon is introduced.  After adding
the photon fragmentation contribution this sample becomes independent of
$y_0^{\gamma}$, i.e.\ also $y_0^{\gamma}$ has the status of a technical
cut like $y_0^J$. In the $\gamma + 1$ parton-level jet event sample the
photon is isolated from $q$ and $\bar{q}$ by requiring a non-zero
transverse momentum of the photon in the $\gamma ^{*}p$ center-of-mass
frame. The two technical cuts $y_0^J$ and $y_0^{\gamma}$ serve only to
distinguish the phase space regions, where the integrations have been
done analytically with arbitrary dimensions from those where they have
been done numerically in four dimensions.  These two parameters must be
chosen sufficiently small to justify the neglect of terms proportional
to $y_0^J$ and $y_0^{\gamma}$, respectively in the analytical part of
the calculations.  We found that $y_0^J=10^{-4}$ and $y_0^{\gamma} =
10^{-5}$ is sufficient to fulfill this requirement.
 
In order to comply with the jet definitions in future analyses of
experimental measurements the partons and the photon in the $\gamma + 2$
parton-level jet sample are recombined to $\gamma + (1+1)$-jets using
the cone algorithm of the Snowmass convention \cite{snowmass}. In this
recombination scheme the photon is treated like any other parton
(so-called democratic algorithm). In the $\gamma^{*}p$ center-of-mass
frame, two partons $i$ and $j$ are combined into a jet $J$ if they obey
the cone constraint $R_{i,J} < R$, where
\begin{equation}
  R_{i,J} = \sqrt{(\eta_i-\eta_J)^2+(\phi_i-\phi_J)^2}.
\end{equation}
$\eta_J$ and $\phi_J$ are the rapidity and azimuthal angle of the
recombined jet. These variables are obtained by taking the averages of
the corresponding variables of the recombined partons $i$ and $j$
multiplied with their $p_T$ values. The $p_T$ of jet $J$ is the sum of
$p_{T,i}$ and $p_{T,j}$. We choose $R = 1$.  In some cases, an ambiguity
may occur, when two partons $i$ and $j$ qualify both as two individual
jets $i$ and $j$ and as a combined jet $J$. In this case we count only
the combined jet $J$ to avoid double counting. The rapidity is always
defined positive in the direction of the proton remnant momentum.  The
azimuthal angle is defined with respect to the scattering plane given by
the momentum of the beam and the scattered electron. One of the
recombined jets may be the photon jet. To qualify a jet as a photon jet
we restrict the hadronic energy in this jet by requiring
\begin{equation} 
z_{\gamma} = \frac{p_{T,\gamma}}{p_{T,\gamma}+p_{T,had}} = 
1 - \epsilon_{had} > z_{cut}. 
\end{equation}

$p_{T,\gamma}$ and $p_{T,had}$ denote the transverse momenta of the
photon and the parton producing hadrons in this jet, respectively. For
our predictions we choose $\epsilon_{had} \leq \epsilon^0_{had} = 0.1$
\cite{2}.  Eventually the parameter $\epsilon^0_{had}$ (or $z_{cut}$)
must be chosen in accordance with the experimental analysis. Our results
depend also on the choice of the quark-to-photon fragmentation function.
The factorization scale dependent quark-to-photon FF of $O(\alpha)$ is 
taken from \cite{13}.  It is the sum of two contributions, the solution 
of the evolution equation at this order and an initial FF at some 
initial scale $\mu_0$.  The initial FF function and initial scale have 
been fitted to the ALEPH $\gamma + 1$-jet data \cite{8}. The 
factorization scale dependent quark-to-photon fragmentation function 
also gives a good description of the inclusive photon distribution as 
measured by OPAL \cite{9}.

With these definitions it is clear that in NLO the final state may
consist of two or three jets, where one jet is always a photon jet. The
three-jet sample, equivalent to $\gamma + (2+1)$-jets in the notation of
the previous sections, consists of all $\gamma + (2+1)$ parton
level jets, which do not fulfill the cone constraint (10).

\begin{figure}[htbp] 
\unitlength 1mm
\begin{picture}(160,100)
\put(-1,20){\epsfig{file=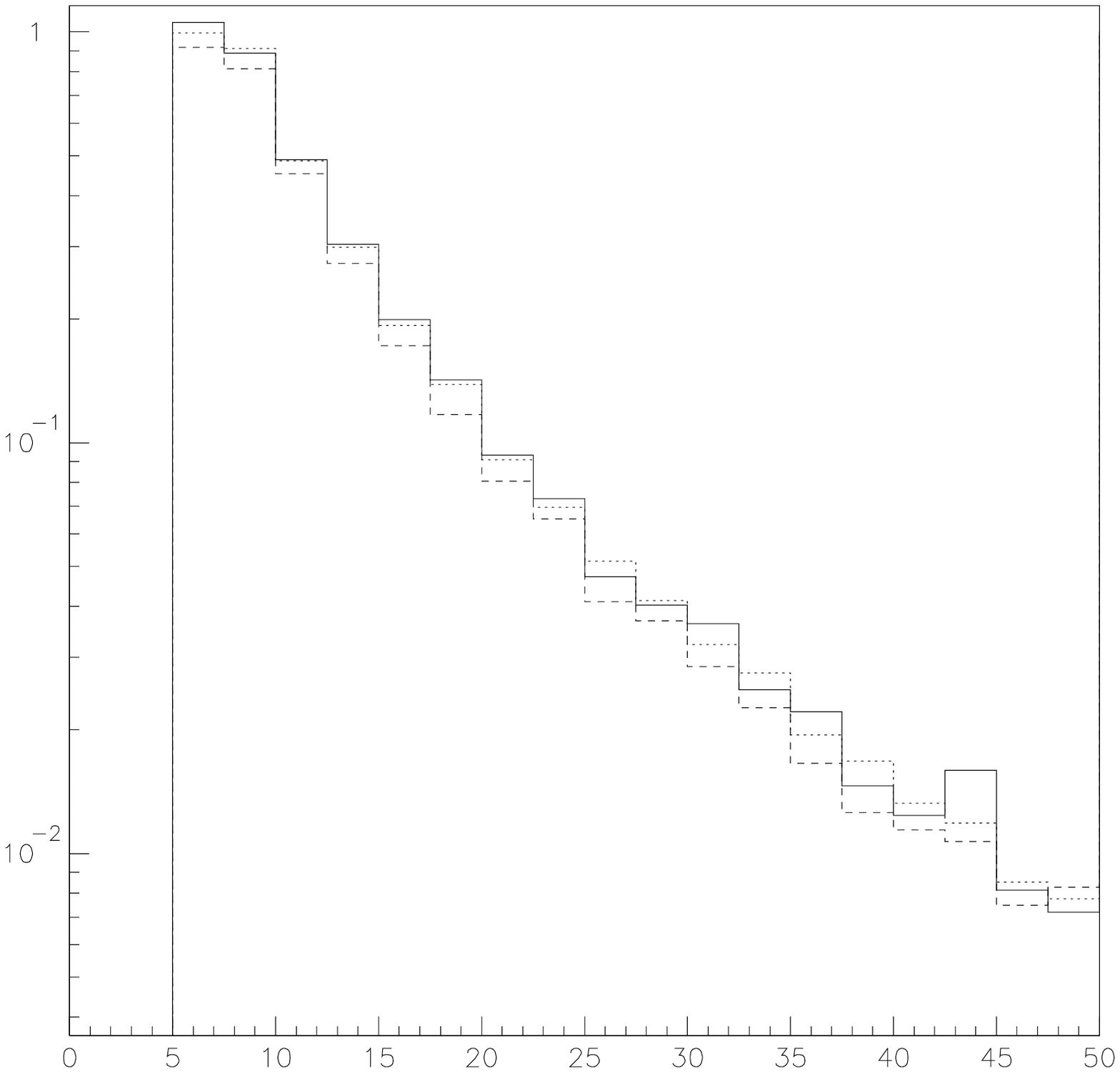,height=8cm,width=8cm,%
                   bbllx=40,bblly=180,bburx=538,bbury=655,clip=}}
\put(45,86){$\displaystyle\frac{d\sigma}{dp_{T,\gamma}}$[pb/GeV]}
\put(61,17){\footnotesize $p_T$[GeV]}
\put(38,15){(a)}
\put(82,20){\epsfig{file=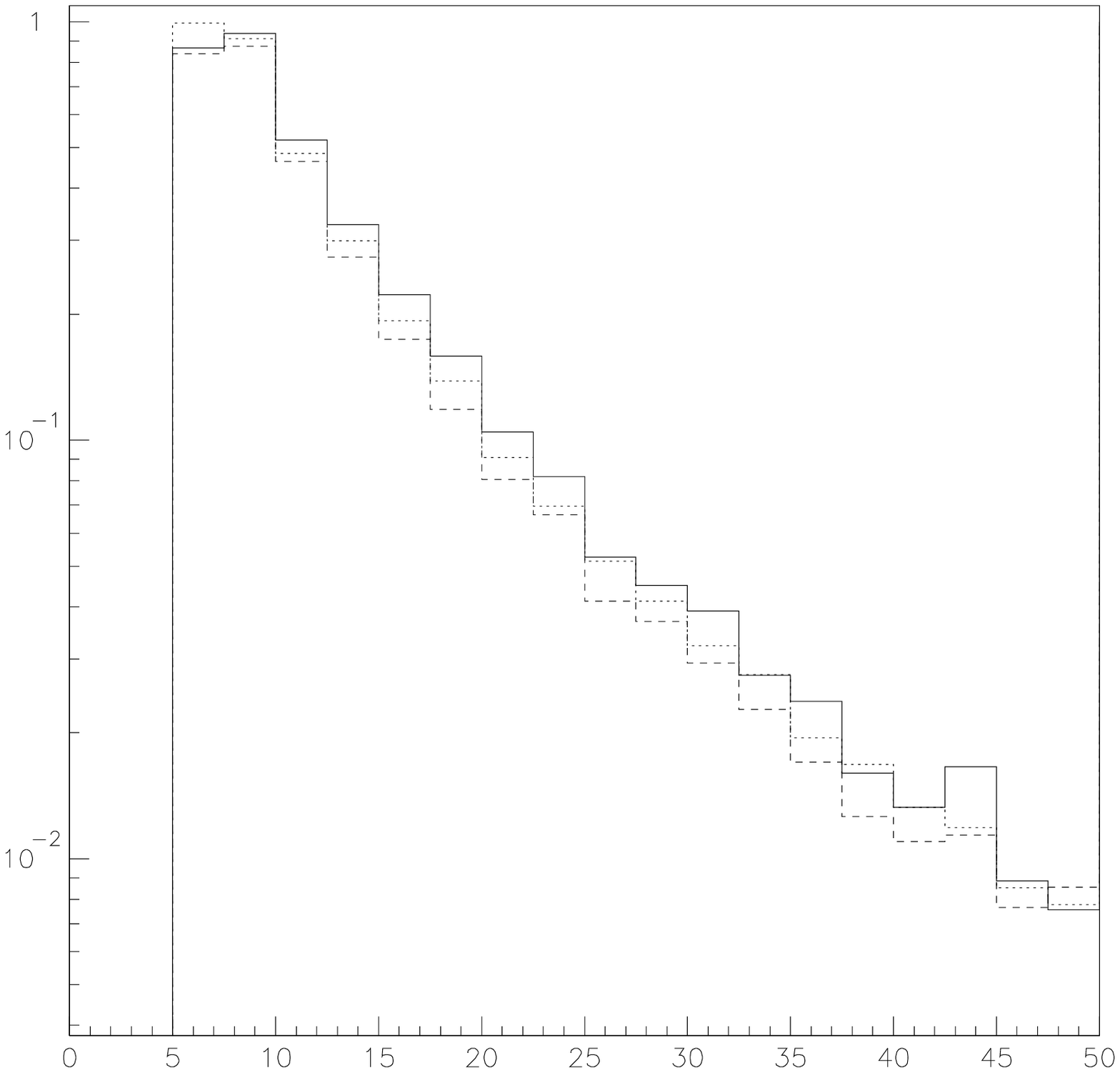,height=8cm,width=8cm,%
                   bbllx=40,bblly=180,bburx=538,bbury=655,clip=}}
\put(126,86){$\displaystyle\frac{d\sigma}{dp_{T,J}}$[pb/GeV]}
\put(144,17){\footnotesize $p_T$[GeV]}
\put(121,15){(b)}
\put(0,6){\parbox[t]{16cm}{\sloppy Figure 1: $p_T$ distributions of 
    photon (a) and jet with largest $p_T$ (b) for LO (dotted line), NLO 
    $\gamma +(1+1)$-jets (dashed line) and NLO $\gamma +(1+1)$- and
    $\gamma +(2+1)$-jets (full line).  }}
\end{picture}
\end{figure}

\begin{figure}[htbp] 
\unitlength 1mm
\begin{picture}(160,100)
\put(-1,20){\epsfig{file=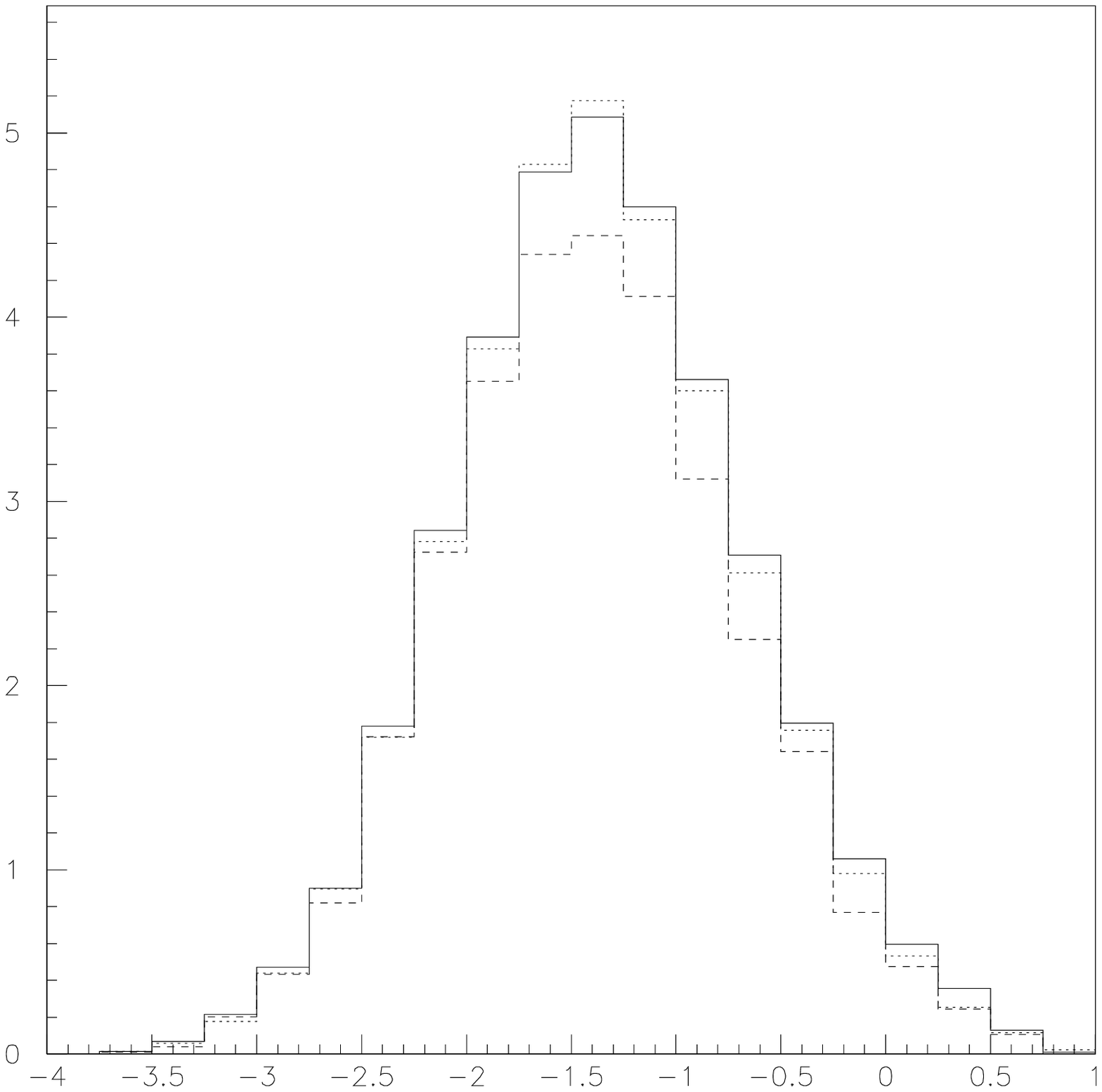,height=8cm,width=8cm,%
                   bbllx=40,bblly=180,bburx=538,bbury=655,clip=}}
\put(11,86){$\displaystyle\frac{d\sigma}{d\eta_{\gamma}}$[pb]}
\put(38,15){(a)}
\put(82,20){\epsfig{file=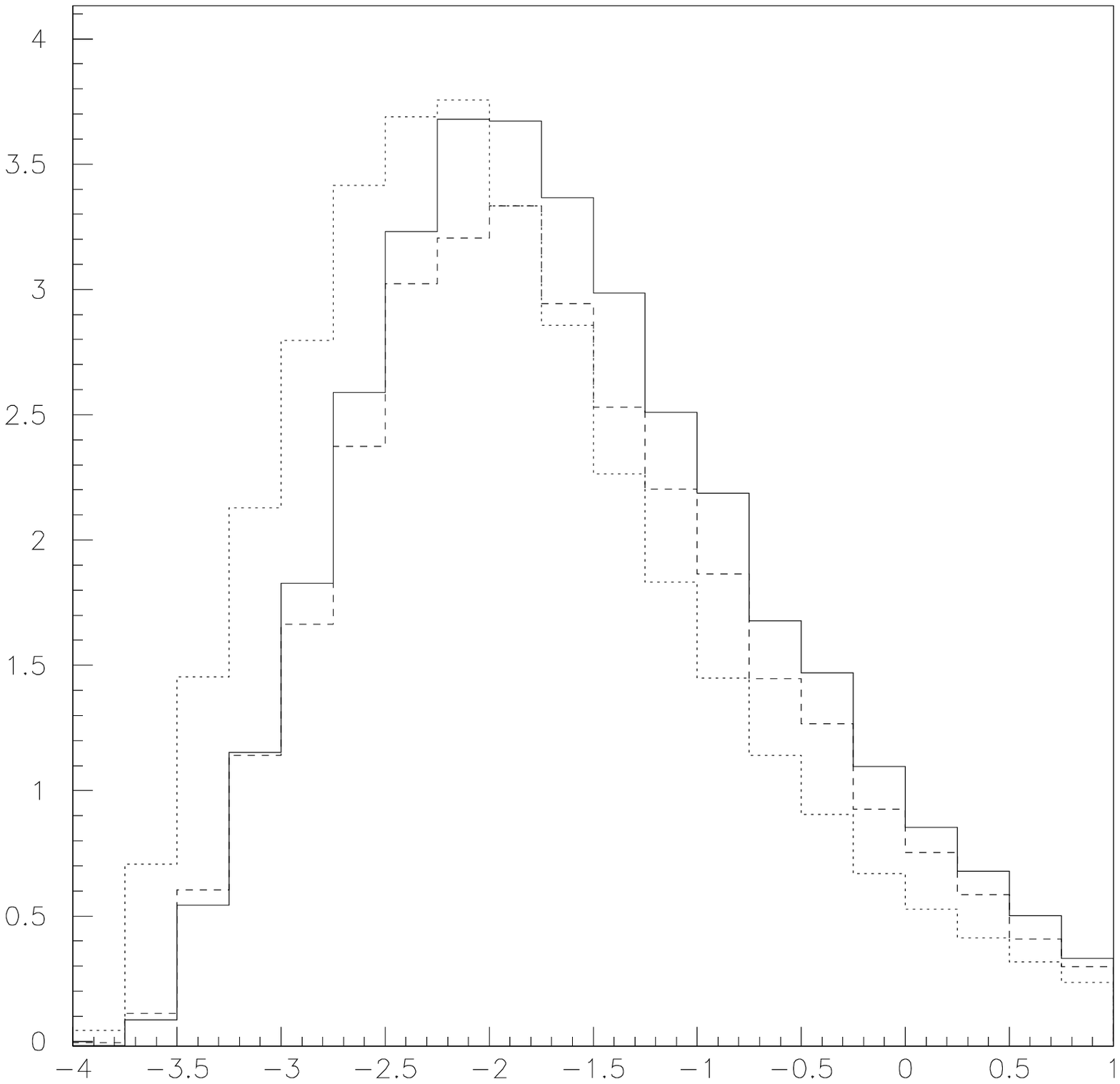,height=8cm,width=8cm,%
                   bbllx=40,bblly=180,bburx=538,bbury=655,clip=}}
\put(138,86){$\displaystyle\frac{d\sigma}{d\eta_J}$[pb]}
\put(121,15){(b)}
\put(0,6){\parbox[t]{16cm}{\sloppy Figure 2: $\eta$ distributions of
    photon (a) and jet with largest $p_T$ (b) with labeling of curves  
    as in Fig.\ 1. }}
\end{picture}
\end{figure}

In Fig.\ 1 and 2 we show our results for the $p_T$ and $\eta$ dependence
of the cross sections, $d\sigma/dp_T$ and $d\sigma/d\eta$, concerning
the photon and the jet with the largest $p_T$. In each figure we have
plotted three curves, (i) the cross section in LO (dotted curve), which
has only $\gamma + (1+1)$-jets and is independent of the jet defining
parameters $R$ and $\epsilon^0_{had}$, (ii) the NLO cross section for
$\gamma + (1+1)$-jets (dashed curve) and (iii) the sum of the NLO cross
sections for $\gamma + (1+1)$-jets and $\gamma + (2+1)$-jets (full
curve).  Specifically, in Fig.\ 1a we present $d\sigma/dp_{T,\gamma}$,
the transverse momentum dependence of the three cross sections (i), (ii)
and (iii) for $p_{T,\gamma} \geq 5~GeV$. All other variables, in
particular $\eta_J$, $\eta_{\gamma}$ and $p_{T,J}$, are integrated over
the kinematically allowed ranges. We see that all three cross section
have a similar shape. The sum of the $\gamma + (1+1)$- and $\gamma +
(2+1)$-jets cross section is only slightly larger than the $\gamma +
(1+1)$-jets cross section. Both cross sections do not differ very much
from the LO cross section indicating that the NLO corrections are not
very large. Of course, this is a consequence of our choice for the cone
radius $R$. In Fig.\ 1b the plot of $d\sigma/dp_{T,J}$ for the jet with
the largest $p_T$ is shown. The qualitative behaviour of the three cross
sections (i) - (iii) is similar as in Fig.\ 1a. For the $\eta$
distributions we integrate over $p_{T,\gamma} \geq 5~GeV$ and $p_{T,J}
\geq 6~GeV$. The choice of two different values of minimal $p_T$'s for
the photon and the jet is necessary to avoid the otherwise present
infrared sensitivity of the NLO predictions. This sensitivity is known
from similar calculations of dijet cross sections in $ep$ collisions
\cite{18} and must be avoided.  The cross section
$d\sigma/d\eta_{\gamma}$ is plotted in Fig.\ 2a, again for the three
cases (i), (ii) and (iii). The shapes of the three curves are similar.
Here we have integrated over the full kinematic range of the variable
$\eta_J$. Figure 2b contains the predictions for $d\sigma/d\eta_J$,
where $\eta_J$ is the rapidity of the jet with the largest $p_T$.  In
Fig.\ 2b we observe that the full curve, which represents
$d\sigma/d\eta_J$ for the sum of the two jet cross sections, is shifted
somewhat more to $\eta_J > 0$ compared to the NLO $\gamma + (1+1)$-jets
cross section.  The LO cross section (dotted curve) peaks more in the
backward direction than the other two. Compared to
$d\sigma/d\eta_{\gamma}$, shown in Fig.\ 2a the $\eta_J$ distribution
for the jet peaks at somewhat smaller $\eta_J$.

\begin{table}[ttt]
\renewcommand{\arraystretch}{1.3}
\caption{Contributions to the cross sections (in pb) for $\gamma
  +(1+1)$- and $\gamma +(2+1)$-jet production (see text).}  
\begin{center}
\begin{tabular}{lrr} \hline
 Contribution   & $\gamma +(1+1)$-jets & $\gamma +(2+1)$-jets \\ \hline
 \hline 
 LO             &   $8.561 \pm 0.022$  &                      \\ \hline
 S              & $-32.164 \pm 0.026$  &                      \\ \hline
 R($q, \bar{q}$)&  $24.843 \pm 0.034$  & $0.6349 \pm 0.0008$  \\ \hline
 R($g$)         &   $3.717 \pm 0.005$  & $0.3239 \pm 0.0007$  \\ \hline
 F              &   $3.429 \pm 0.022$  &                      \\ \hline
 D              &  $-0.599 \pm 0.001$  &                      \\ \hline
 sum            &   $7.787 \pm 0.053$  & $0.9588 \pm 0.0010$  \\ \hline
\end{tabular}
\end{center}
\end{table}

For completeness we also give the results for the various cross sections
in Table 1, separated into the contributions S, R (for incoming quarks
or antiquarks and incoming gluons), F as described in the previous
section and the fragmentation contribution denoted by D as defined in 
Eq.\ (5). About $12\,\%$ of the total NLO cross section is due to 
$\gamma +(2+1)$-jet final states. The $\gamma +(1+1)$-jets cross section 
is reduced by about $9\,\%$ by NLO corrections.

%
\section{Concluding Remarks} 

We have presented a NLO calculation for the production of photons
accompanied by jets in deep inelastic electron proton scattering,
taking into account the contribution from quark-to-photon
fragmentation. This improves a previous work which suffered from the
presence of parton-level cutoff parameters. The present consistent 
treatment allows for a direct comparison of our theoretical predictions
with experimental measurements without being sensitive to uncertainties
from unphysical cutoff parameters. 

We expect that the measurement of photon plus jet production at HERA
will contribute to testing perturbative QCD. Moreover, our results add
another piece to the set of NLO predictions of the standard model needed
in searches for new physics. The calculation covers the range of large
$Q^2$ up to several $10^3~GeV^2$. At even larger momentum transfers
additional contributions from $Z$ exchange become as important as pure
$\gamma$ exchange to which the present work was restricted.

\subsection*{Acknowledgements}

A.\ G.\ would like to thank A.\ Wagner for financial support during her
stay at DESY where part of this work has been carried out.


\end{document}